\documentclass[prd,twocolumn,nofootinbib,aps,showpacs]{revtex4}

\usepackage{amsmath}
\usepackage{amssymb}
\usepackage{graphicx}
\usepackage{color}
\usepackage{bm}
\usepackage{times}
\usepackage{pdfpages}
\usepackage{hyperref}
\hypersetup{
  colorlinks=true,
  citecolor=blue,
  linkcolor=blue,
  urlcolor=blue}

\everymath{\displaystyle}

\usepackage{epsfig}
\usepackage{dcolumn}
\usepackage{float}

\newcommand{\dcp}{\delta_{CP}}
\newcommand{\nova}{NO$\nu$A\ }

\begin{document}


\title{Reason for T2K to run in dominant neutrino mode for detecting CP violation
}

\author{Monojit Ghosh}
\email[Email Address: ]{monojit@prl.res.in}
\affiliation{
Physical Research Laboratory, Navrangpura,
Ahmedabad 380 009, India}

\begin{abstract}
The long-baseline experiment T2K in Japan has collected data in the neutrino mode corresponding to
an exposure of $6.57 \times 10^{20}$ POT (Protons on Target) and currently it is running in the antineutrino mode.
The main aim of the antineutrino run is to measure the leptonic phase $\dcp$ 
which may help to understand the matter-antimatter asymmetry of the universe. 
In this work we show that in T2K,
antineutrinos are required only for removing 
the wrong octant solutions which in turn improves the CP sensitivity. 
If however the octant is known then pure neutrino run is capable of giving 
the maximum CP sensitivity.
If we divide the total true parameter space into eight sets, then we find that
T2K antineutrino run helps in improving the CP sensitivity for only two sets while for
the remaining six combinations pure neutrino run gives the best CP sensitivity. Thus if the neutrino run is replaced by the antineutrino run
then it causes a reduction in the CP sensitivity in 75\% of the true parameter space due to lesser statistics.
Thus it is worthwhile to study
if the T2K antineutrino run can be reduced by the antineutrino runs of the other experiments, so that T2K can run in dominant neutrino mode
to extract the best CP sensitivity. 
In this work we explore
the possibility of the antineutrino component of NO$\nu$A and the atmospheric neutrino experiment ICAL@INO for compensating the antineutrino run of T2K.

\end{abstract}



\maketitle

\section{Introduction}

The discovery of non zero $\theta_{13}$, from the recent reactor experiments Daya Bay \cite{Zhang:2015fya}, Double Chooz \cite{Crespo-Anadon:2014dea},
and RENO \cite{Kim:2014rfa}, 
bring the remaining unknowns in neutrino oscillation physics
within a reach. 
Among the six parameters that describe the phenomenon of neutrino oscillation 
(i.e., three mixing angles $\theta_{12}$, $\theta_{13}$ and $\theta_{23}$, two mass squared
differences $\Delta_{21} (m_2^2 - m_1^2)$ and $\Delta_{31} (m_3^2 - m_1^2) $, and the Dirac type CP phase $\dcp$) 
, the still undetermined parameters are:
i) the neutrino mass orderings which tell if neutrinos are normal hierarchical (NH) i.e., $\Delta_{31}$ is +ve or inverted hierarchical (IH) i.e.,  $\Delta_{31}$
is -ve,
ii) the octant of $\theta_{23}$
which defines if $\theta_{23}$ belongs to the lower octant (LO) i.e., $< 45$ or higher octant (HO) i.e., $> 45$
and finally
iii) the exact value of $\dcp$.
There are several current and future generation oscillation experiments
which are dedicated to measure these remaining unknowns. Among these, the long-baseline experiments like T2K \cite{Abe:2011sj}, 
\nova \cite{Ayres:2004js}, DUNE \cite{Adams:2013qkq}, 
ESS$\nu$SB \cite{Baussan:2013zcy} will
use neutrinos from the accelerator to observe the flavour transition, on the other hand different atmospheric neutrino
experiments like Super-Kamiokande \cite{Wendell:2010md}, Hyper-Kamiokande \cite{Abe:2011ts}, ICAL@INO \cite{inowebsite}, 
PINGU \cite{Aartsen:2014oha} will make use of the neutrinos coming
from the Earth's atmosphere to detect neutrino oscillation. 
It is to be noted that the optimal design of any current experiment requires 
the understanding of the parameter space to which the experiment is sensitive to and simultaneously
how the data from other experiments can help in improving its sensitivity \cite{Ghosh:2013pfa,Ghosh:2014rna}.  

T2K is an ongoing accelerator based long-baseline
experiment in Japan. The neutrino flux is generated 
at the J-PARC site on the East coast of Japan and 
directed to the Super-Kamiokande neutrino detector in the mountains of western Japan.
The narrow band flux peaks at 0.6 GeV which coincides with the T2K appearance channel oscillation maxima ($\nu_\mu \rightarrow \nu_e$ oscillation).
As the baseline of this experiment is only 295 km, it has very limited sensitivity to neutrino mass hierarchy and octant but on the other hand can provide good
CP sensitivity. T2K has already given the first set of data providing a hint at $\dcp=-90^\circ$ \cite{Abe:2015awa}\footnote{The recent results from the \nova
also prefer $\dcp$ near to $-90^\circ$ \cite{Bian:2015opa}.}.
This hint comes from a run of $7.8\%$ of the proposed total exposure in the neutrino mode \cite{Abe:2013hdq}.
Currently T2K is taking data in antineutrino mode to establish CP violation in leptonic sector on a firm footing. 
At this stage it is important to ask if T2K should run in a dominant neutrino mode or it should run in equal neutrino-antineutrino mode.
To answer this question one needs to understand the underlying physics of 
how antineutrinos actually help in improving the CP sensitivity of a particular experiment.
The determination of
$\dcp$ in long-baseline experiments
is constrained by the parameter degeneracy \cite{degen,Coloma:2014kca,Machado:2013kya,Ghosh:2015ena,Minakata:2013hgk}.
In particular, the limited hierarchy and octant sensitivity of T2K,
give rise to hierarchy-$\dcp$ degeneracy (degeneracy of type I)
and octant-$\dcp$ degeneracy (degeneracy of type II). 
The behaviour of type I degeneracy is similar in the neutrino and antineutrino oscillation probability \cite{Prakash:2012az} but
the type II degeneracy behaves differently in neutrinos and antineutrinos \cite{Agarwalla:2013ju}. So while determining $\dcp$, 
addition of antineutrinos over neutrinos can help in
removing the wrong octant solutions but not  the 
wrong hierarchy solutions.
Apart from the synergy described above, the role played by antineutrinos also depends on the baseline and flux profile of a particular experiment. 
In this work we will quantify this point
by showing that if octant is known then antineutrino run is not at all required for any true combination for T2K. This implies that because of 
the T2K baseline and the flux
profile there is no other contribution of the antineutrinos in T2K apart from removing the type II degeneracy. On the contrary we show that
the situation is different in \nova.

\nova is also an accelerator based long-baseline experiment at Fermilab which has a baseline of 812 km. At this baseline
the first oscillation maximum occurs at 1.6 GeV whereas the NUMI off axis flux peaks at slightly higher energy which is 2 GeV. 
\nova has already given its first results \cite{nova_recent} and
it is planned to run in equal neutrino-antineutrino mode. 
In our study we find that there exists a
synergy between neutrinos and antineutrinos in \nova which helps in improving the CP sensitivity even when
the octant is known. This difference between T2K and \nova for measuring $\dcp$,
motivates us to study if the antineutrino run of T2K can be compensated by
antineutrino runs of other experiments and \nova in particular. 
If T2K runs in dominant neutrino mode then the measurement of $\dcp$ for the true parameter space that does not suffer from
any octant degeneracy can be done without compromising on the 
statistics which  gets reduced if one adds equal amount of antineutrinos
as that of neutrinos. For the parameter space where 
antineutrino run is required the antineutrino component of the other 
experiments can provide the required sensitivity. 
In this way T2K, with the help of other
experiments can contribute to getting the first signature of $\dcp$ using 
its maximum statistics by running in dominant neutrino mode.
This point has been also raised in \cite{Evslin:2015pya} in the context of T2HK \cite{Abe:2015zbg}, which considers antineutrinos from a muon decay at rest (DAR) source.
In this work we have focused on the T2K experiment using the existing/funded facilities.

We have considered the options of adding \nova and the 
atmospheric neutrino experiment ICAL@INO to T2K.
The India-based Neutrino Observatory (INO) Project is a multi-institutional effort aimed at building a world-class underground 
laboratory for detecting muon events arising from the flavour oscillation of the atmospheric neutrinos. It will use Iron Calorimeter (ICAL) detector, 
consisting of a 50 kiloton magnetized iron plates arranged in stacks with gaps in between where Resistive Plate Chambers (RPCs)
would be inserted as active detectors.The magnetic field will allow to analyze the neutrino and antineutrino events separately.
As sensitivity of the atmospheric neutrino gets contribution from both neutrinos and antineutrinos, 
it is also relevant to study the combined sensitivity of atmospheric data
 and T2K.  

For numerical simulation  we have considered a total
exposure of $8\times 10^{21}$ POT for T2K. We have divided the exposures in the units of $10^{21}$ POT and generated our results for different neutrino-antineutrino
exposures. For \nova we have taken an equal 3 years run in both neutrino and antineutrino mode unless otherwise stated. 
The T2K and \nova results are simulated using GLoBES \cite{globes,globes1,globes2,globes3,globes4,globes5,globes6,globes7,globes8,globes9}. 
For ICAL we have used an in-house code with a fixed resolution
of $10\%$ in energy and $10^\circ$ for the direction and generated the data for a total exposure of 500 kt-yr. 

The plan of the paper goes as follows. In Section \ref{sec2}, we briefly discuss the degenerate parameter space of T2K in terms of oscillation probability and also discuss
the role of antineutrinos in solving these degeneracies at the probability level. In Section \ref{sec3}, 
we will quantify the role of antineutrinos in T2K for determining the leptonic CP phase $\dcp$ in terms of CP violation discovery (CPV).
We will also identify the parameter space where the antineutrino runs are necessary. In Section \ref{sec4},
we will study the role of antineutrinos in \nova to discover CPV. Then in Section \ref{sec5},
we study the possibility of reduce the antineutrino runs of T2K by the
addition of \nova and ICAL data and try to find the optimal configuration for T2K. Then finally in Section \ref{sec6}, we will summarize our results and then conclude. 

\section{The degenerate parameter space in T2K probability}
\label{sec2}

\begin{figure*}
\begin{tabular}{rcl}
\epsfig{file=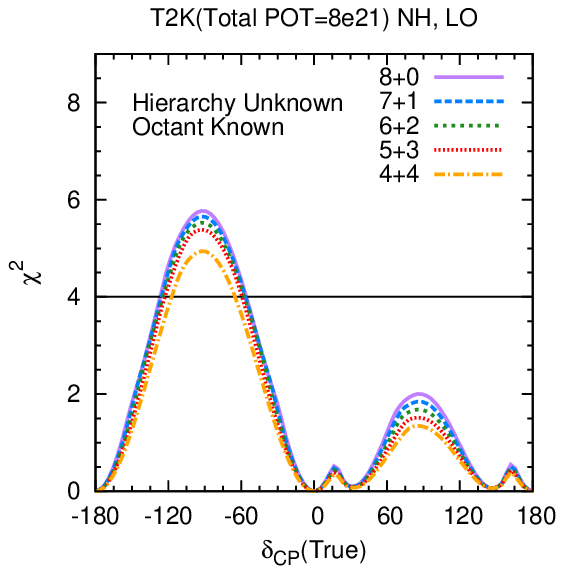, width=0.4\textwidth, bbllx=89, bblly=50, bburx=260, bbury=255,clip=}
\epsfig{file=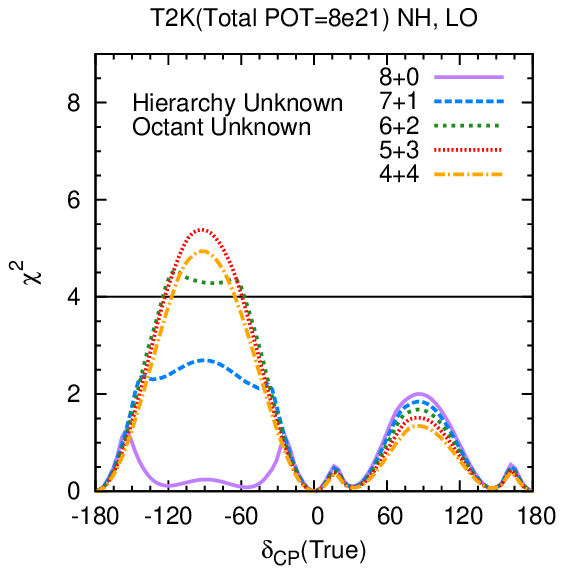, width=0.4\textwidth, bbllx=89, bblly=50, bburx=260, bbury=255,clip=} \\
\epsfig{file=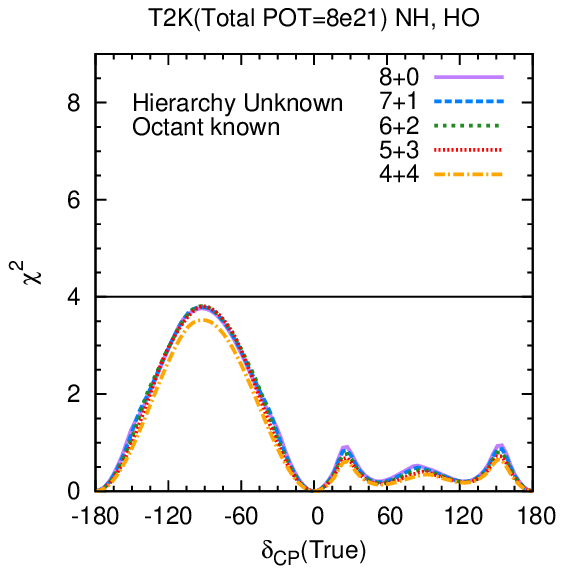, width=0.4\textwidth, bbllx=86, bblly=50, bburx=260, bbury=255,clip=}
\epsfig{file=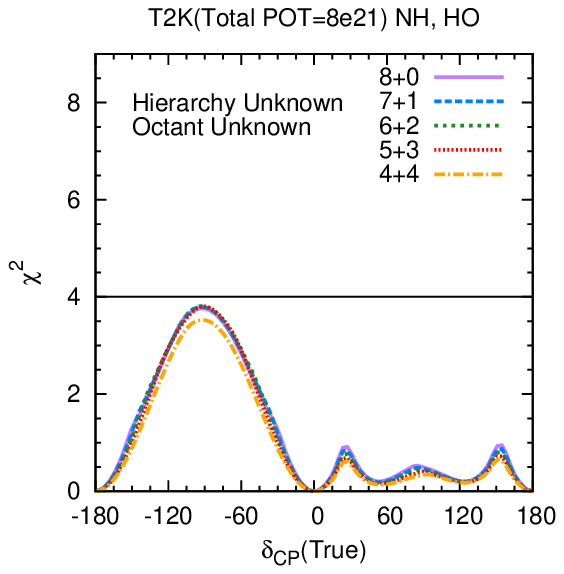, width=0.4\textwidth, bbllx=86, bblly=50, bburx=260, bbury=255,clip=}
\end{tabular}
\caption{CPV $\chi^2$ vs $\dcp$(True) of T2K for NH with $\theta_{23}$(true) $= 39^\circ (51^\circ)$ for LO (HO) and $\Delta_{31}$(true) = $2.4 \times 10^{-3}$.}
\label{NH}
\end{figure*}

For T2K, the CP sensitivity comes from the appearance channel $P_{\mu e}$. At this baseline, the constant density approximation holds good and 
 the probability formula for neutrinos in terms of the small parameters $s_{13}(=\sin\theta_{13})$ and $\alpha(=\frac{\Delta_{21}}{\Delta{31}})$ can be expressed as 
\cite{akhmedov}
\begin{widetext}
\begin{eqnarray}
P_{\mu e} &=& 4 s^{2}_{13}s^{2}_{23}\frac{\sin^{2} (A-1)\Delta}{(A-1)^2} 
     +\alpha s_{13} \sin 2\theta_{12}  \sin 2\theta_{23}\cos(\Delta+\delta_{cp})
     \frac{\sin (A-1)\Delta}{(A-1)}\frac{\sin A\Delta}{A} + {\cal{O}}(\alpha^2) ~,
\label{p_mu_e}
\end{eqnarray}
where $ s_{ij}(c_{ij})=\sin \theta_{ij}(\cos \theta_{ij}) $, $A = 0.000076 \times \rho \times E/\Delta_{31}$, $\Delta = \Delta_{31}L/4E$. $E$ is the energy of the neutrino
in GeV, $L$ is the baseline in km and $\rho$ is the matter density in units of gm/cc. $\Delta_{31}$ and $A$ is +ve for NH and -ve for $IH$. The expression
for antineutrinos can be obtained by replacing $\dcp \rightarrow -\dcp$ and $A \rightarrow -A$.
\end{widetext}
Due to the comparatively smaller baseline, T2K matter oscillation 
maxima coincides
with the vacuum peak and thus $\Delta$ corresponds to $90^\circ$.  As the T2K flux peaks around this same energy where the oscillation maxima peaks,
it will be sufficient to discuss the CP property
of T2K
at this value $\Delta$. 
In this condition, the CP dependent term goes as $\sin\dcp$ and because of this for both neutrinos and antineutrinos $\pm 90^o$
are the maximum separated points in the probability for both the hierarchies.
Below we are listing the degenerate parameter space of T2K which
affects the measurement of $\dcp$:

i) For a given value of $\theta_{23}$ ($\in$ LO or HO), NH having $0^\circ < \dcp < 180^\circ$ (upper half plane or UHP) is 
degenerate with IH having $-180^\circ < \dcp <  0^\circ$ (lower half plane or LHP) i.e., 
$P_{\mu e}(\text{NH},\text{UHP})=P_{\mu e}(\text{IH},\text{LHP})$.
This parameter
space is same for both neutrino and antineutrinos as both the hierarchy sensitive term $A$ and $\dcp$ flips its sign. This is the degeneracy of type I also
known as hierarchy-$\dcp$ degeneracy.

ii) For a given hierarchy ($\in$ NH or IH), LO having  $-180^\circ < \dcp <  0^\circ$ is degenerate with HO having $0^\circ < \dcp < 180^\circ$ in neutrinos. i.e., 
$P_{\mu e}(\text{HO},\text{UHP})=P_{\mu e}(\text{LO},\text{LHP})$.
For antineutrinos the degenerate parameter space can be found out by simply changing the sign of $\dcp$.  This is the degeneracy of type II also
known as octant-$\dcp$ degeneracy.

Now because of the above mentioned degeneracies, CP measurement for a given set of true hierarchy, true octant and true $\dcp$ can be affected by
either wrong hierarchy solutions or wrong octant solutions or both. In the next section we will identify the true parameter space for which 
these wrong solutions are present/absent and the role of antineutrinos.

\begin{figure*}
\begin{tabular}{rcl}
\epsfig{file=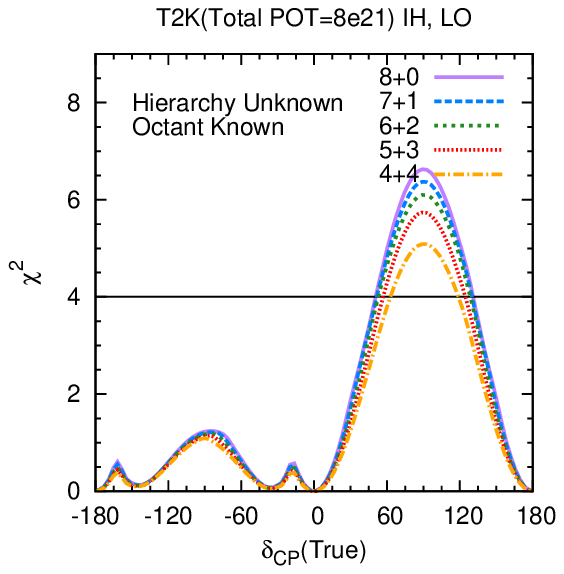, width=0.4\textwidth, bbllx=89, bblly=50, bburx=260, bbury=255,clip=}
\epsfig{file=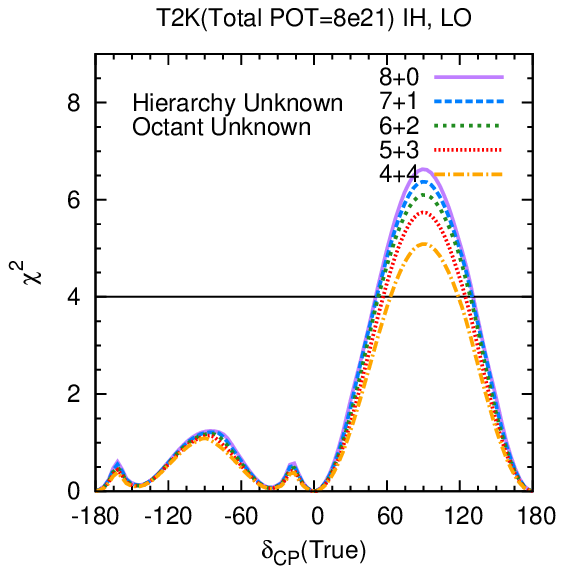, width=0.4\textwidth, bbllx=89, bblly=50, bburx=260, bbury=255,clip=} \\
\epsfig{file=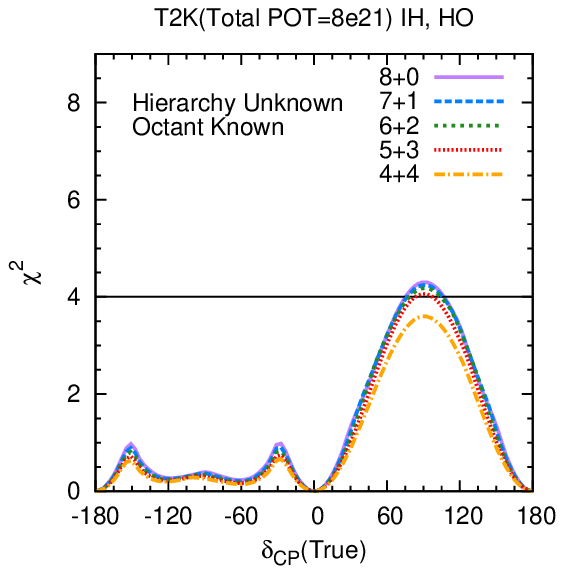, width=0.4\textwidth, bbllx=86, bblly=50, bburx=260, bbury=255,clip=}
\epsfig{file=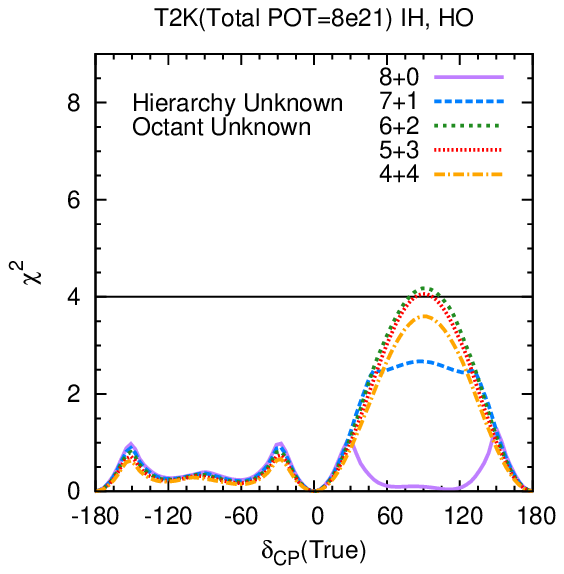, width=0.4\textwidth, bbllx=86, bblly=50, bburx=260, bbury=255,clip=}
\end{tabular}
\caption{ CPV $\chi^2$ vs $\dcp$(True) of T2K for IH with $\Delta_{31}$(true) = $-2.4 \times 10^{-3}$. LO and HO correspond to the same values of $\theta_{23}$ as in Fig \ref{NH}. }
\label{IH}
\end{figure*}

\section{Role of antineutrinos for CPV discovery in T2K: Reason for T2K to run in dominant neutrino mode}
\label{sec3}

\begin{figure*}
\begin{tabular}{rcl}
\epsfig{file=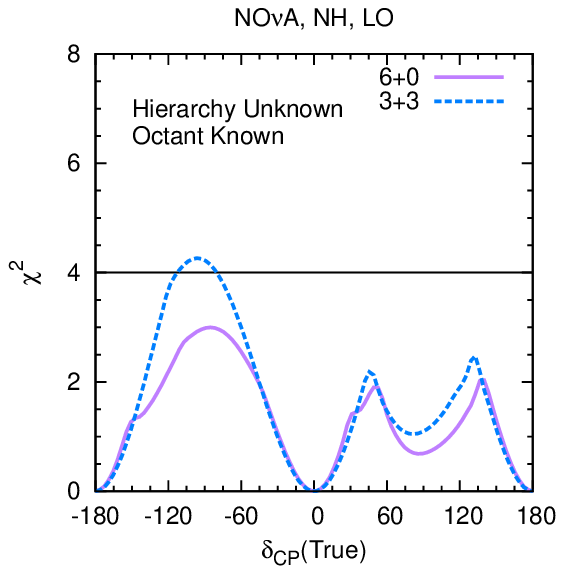, width=0.4\textwidth, bbllx=89, bblly=50, bburx=260, bbury=255,clip=}
\epsfig{file=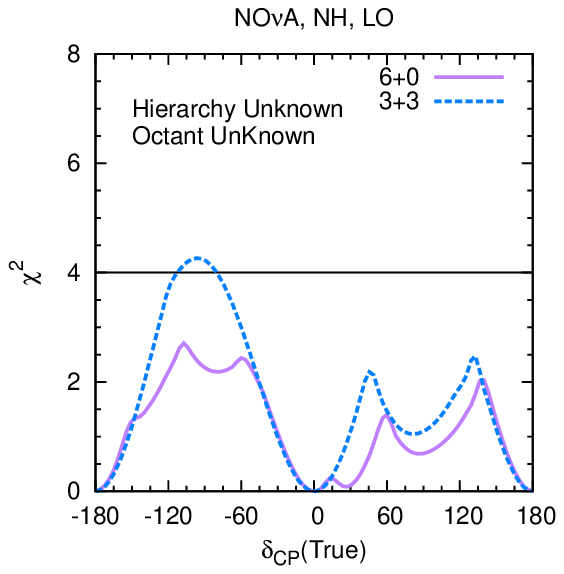, width=0.4\textwidth, bbllx=89, bblly=50, bburx=260, bbury=255,clip=} \\
\end{tabular}
\caption{ CPV $\chi^2$ vs $\dcp$(True) of \nova for NH-LO}
\label{nova}
\end{figure*}

\begin{figure*}
\begin{tabular}{rcl}
\epsfig{file=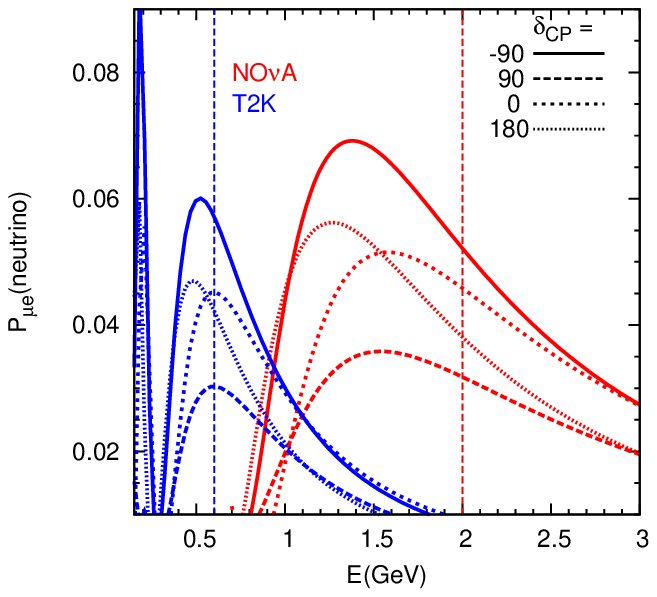, width=0.38\textwidth, bbllx=79, bblly=50, bburx=270, bbury=255,clip=}
\epsfig{file=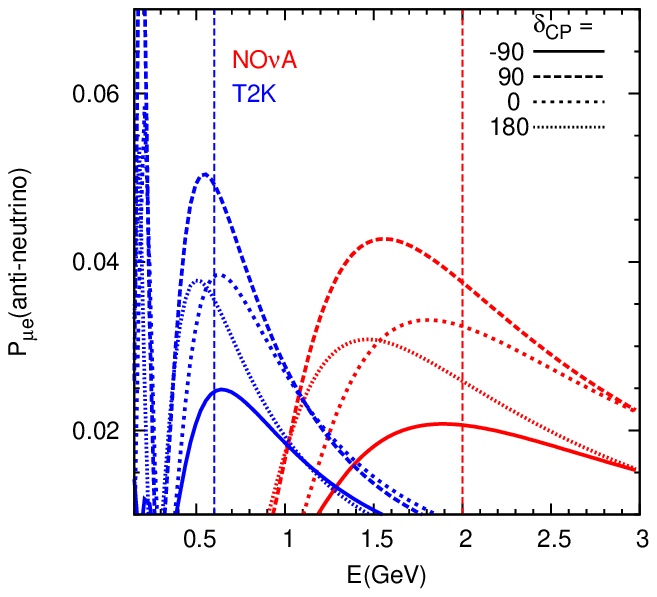, width=0.38\textwidth, bbllx=79, bblly=50, bburx=270, bbury=255,clip=} \\
\end{tabular}
\caption{Probability spectrum for T2K and \nova. The vertical lines correspond to the energy where the respective fluxes peaks}
\label{prob}
\end{figure*}

The CP violation (CPV) discovery potential of an experiment is defined by its capability of distinguishing
the values of $\dcp$ from $0^\circ$ and $180^\circ$.
We calculate the CPV $\chi^2$ by varying $\dcp$ in its full allowed range in 
the true parameter spectrum and keeping its value fixed at $0^\circ$ and $180^\circ$ in the test spectrum.

\begin{figure*}
   \includegraphics[width =\textwidth]{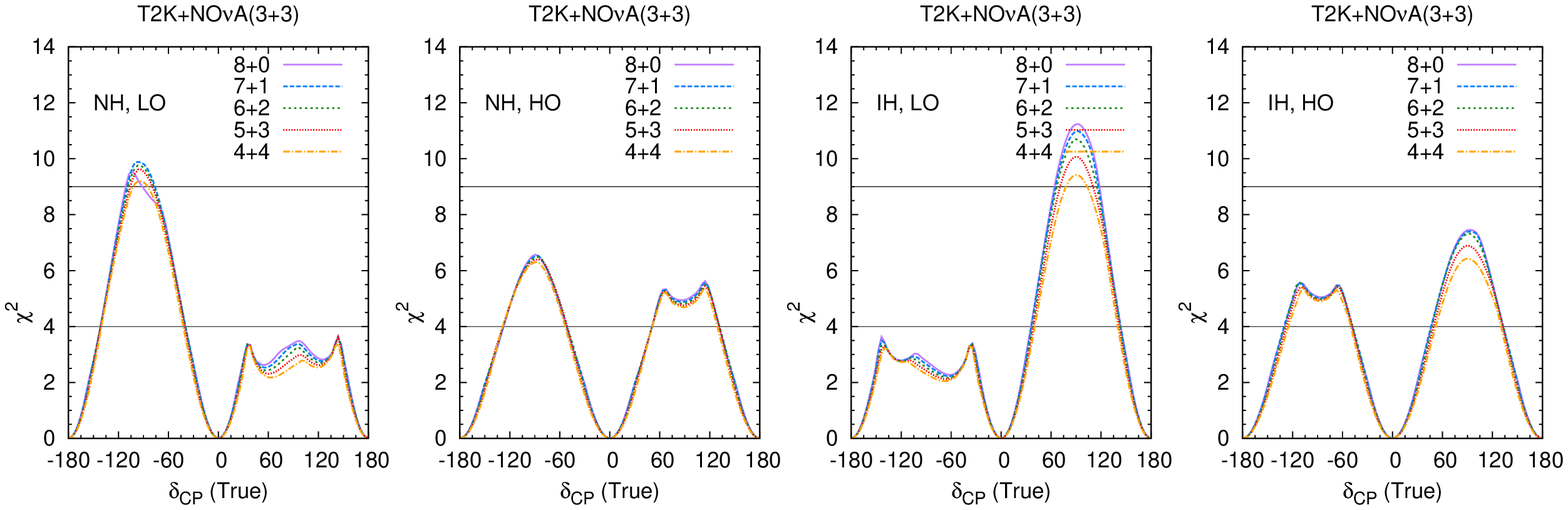}
  \includegraphics[width =\textwidth]{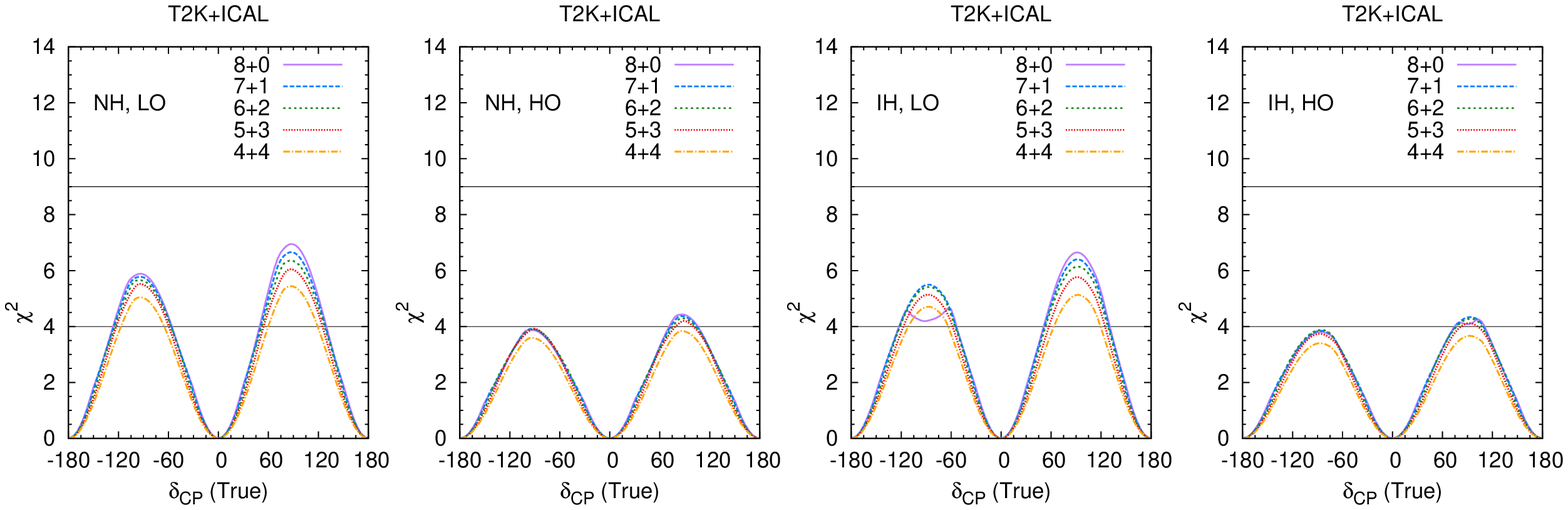}
  \includegraphics[width =\textwidth]{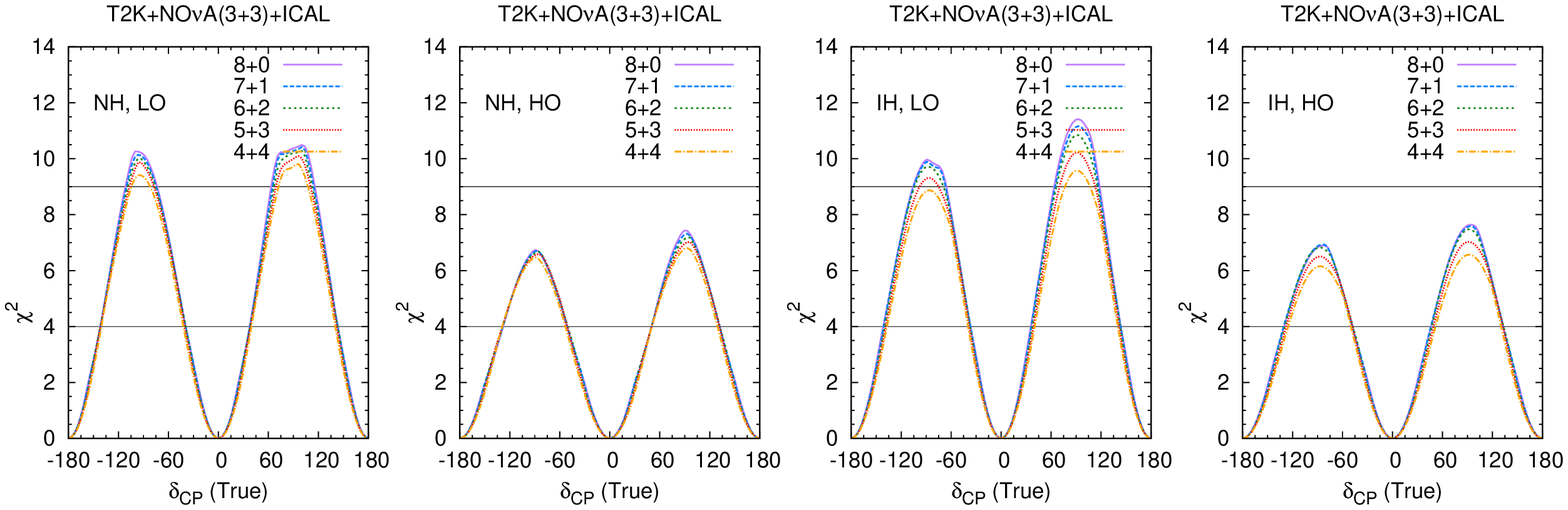}
   \caption{combined CP sensitivity for various anti-neutrino combinations taking unknown hierarchy and unknown octant.}
  \label{ICAL}
   \end{figure*}

In Fig. \ref{NH} we have plotted the CPV discovery $\chi^2$ of T2K by taking true hierarchy as NH.
The upper panels are for true $\theta_{23}$ value of $39^\circ$ which belongs to LO and lower panels correspond to the true $\theta_{23}$ value of 
$51^\circ$ which belongs to HO. 
As we are interested in understanding the role of antineutrinos in detecting CP violation and
we already know that the nature of hierarchy-$\dcp$ degeneracy is same for both neutrinos and antineutrinos, 
we keep the test hierarchy as unknown throughout our analysis. In the left column, the octant has been assumed to be known and in the 
right column the octant is unknown.
To see the role of antineutrinos 
in each panel we have taken five different sets of the neutrino and antineutrino exposures in units of POT. 
In all the panels we can notice a drop in the $\chi^2$ for  $\dcp \in$ UHP. This is because for a given value of $\theta_{23}$, 
NH suffers from hierarchy-$\dcp$ degeneracy for $\dcp \in$ UHP and the $\chi^2$ minima occurs with the wrong hierarchy.
In top left and bottom left panels of Fig. \ref{NH}, we see that
in both the cases (8+0) give the best sensitivity and the sensitivity decreases as the antineutrino run is increased. 
In both the plots the octant has been assumed to be known and this establish the fact that in T2K if octant is known then antineutrinos do not play
any role to improve the CP sensitivity. Moreover addition of antineutrino causes a decrease in the statistics and hence the sensitivity decreases.
In top right panel of Fig. \ref{NH}, when the true octant is LO, we notice that, for $\dcp=-90^\circ$, pure neutrino run of T2K gives 
the worst sensitivity and addition of antineutrinos help in improving the CP sensitivity. For the T2K combination (5+3) the sensitivity becomes maximum and then
the $\chi^2$ falls with the further addition of antineutrinos \cite{Ghosh:2014zea}.
This is because, neutrinos suffer from
octant degeneracy in (LO,LHP) but antineutrinos do not. Thus at this true value, addition of antineutrinos help in improving the CP sensitivity.
But once the degeneracy is removed and the $\chi^2$ minima shifts into the correct octant, further addition of antineutrino causes a decrease in statistics
and hence the sensitivity and as a
result we see (5+3) is better than (4+4). But the situation is different when the true octant is HO (bottom right panel). As neutrinos do not have octant degeneracy at
this point, pure neutrino run gives the maximum sensitivity while addition of antineutrinos only reduces sensitivity.

Now let us discuss the role of antineutrinos when the true hierarchy is inverted. 
In Fig. \ref{IH}, we have plotted the same curves that of Fig. \ref{NH} but assuming the true hierarchy as IH. In all the panels we identify the
drop in the $\chi^2$ sensitivity in the the LHP due the hierarchy-$\dcp$ degeneracy. As in NH, here it is noticed that for the cases when octant is known (top left
and bottom left panels) antineutrinos do not help in the CP sensitivity but when the octant is unknown adding antineutrinos only help in the UHP when the true octant is
HO. This is because for true IH, neutrinos suffers from octant degeneracy at HO-UHP but antineutrinos resolve this degeneracy.
For the other combinations of true octant and CP addition of antineutrinos only reduces sensitivity.

Thus from the above discussion it is clear that
if we divide the total true parameter space into eight sets based on the unknown hierarchy i.e., NH/IH, unknown octant i.e., LO/HO and unknown $\dcp$ i.e., LHP/UHP,
T2K antineutrino run helps in improving the CP sensitivity for only the true combination NH-LO-LHP and IH-HO-UHP. For
the remaining other six combinations, antineutrinos do not have any role to play other than
reducing the sensitivity due to lack of statistics. So if T2K is decided to run in equal neutrino and antineutrino mode, then it will be at the cost of compromising
the CP sensitivity of the 75\% of the total true parameter space. But at the same time it is also true that for the true parameter values for which
antineutrino runs 
are required, pure neutrino run give the worst sensitivity.
To overcome this situation, we suggest that T2K run in dominant neutrino mode, since other experiments will anyway collect antineutrino data.
In this way T2K with the help of other experiments will be able to obtain the best CP sensitivity for 
all values in the true parameter space.

\section{Role of antineutrinos in \nova}
\label{sec4}

Now let us discuss the role of antineutrinos in \nova for detecting CP violation.
To understand the role of anti-neutrinos we have considered \nova(6+0) which corresponds to 6 year running of \nova in the pure neutrino mode and
\nova(3+3) which corresponds to 3 years equal running in both neutrino and antineutrino.

In Fig. \ref{nova} we plot the CPV $\chi^2$ of \nova
for true hierarchy NH and $\theta_{23}=39^\circ$. 
For this true combination, T2K requires antineutrino run in the LHP when octant is unknown but do not need the same when octant is known.
But on the other hand for \nova, we can see that CP sensitivity of \nova(3+3) is better than \nova(6+0) even when octant is known. This implies that
for \nova, antineutrinos not only removes the wrong octant solutions but also gives an extra edge to improve the overall determination of $\dcp$.
This difference between T2K and \nova can be easily explained from Fig. \ref{prob} where we plotted the probability spectrum for both the experiments for
$\theta_{23}=39^\circ$.

Left panel is for neutrinos and the right panel is for antineutrinos. The red curves correspond to \nova and blue curves correspond to T2K. In both the plots
the vertical lines indicate the energy where the flux peaks. The CPV sensitivity of $\dcp=\pm90^\circ$ is determined by the distance between the $\dcp=0^\circ$, 
$180^\circ$ and $\dcp=\pm 90^\circ$ curves. Due to the off-axis configuration of the experiments, it is relevant to look only at the energies where the flux peak.
For T2K we see that in both neutrino and antineutrino, the $\dcp=0^\circ$, $180^\circ$ curves are equidistant from $\dcp=\pm 90^\circ$. This is because for T2K,
the flux peaks at energy very close to the oscillation maxima. At this limit $\cos(\Delta+\dcp)$ goes as $\sin\dcp$ which gives the same probability 
for $\dcp=0^\circ$ and $180^\circ$.
For this reason the sensitivity at any true value of $\dcp$ is same in both neutrinos and
antineutrinos. 
But for \nova as the flux peaks at slightly different energy than the $P_{\mu e}$ oscillation maxima, in neutrinos $-90^\circ$ is closer to $0^\circ$ and in
antineutrinos it is closer to $180^\circ$. Due to this opposite nature in the probability,
the combined sensitivity of neutrinos and antineutrinos give better result as 
compared to only neutrinos.
Though we show our results for only $\theta_{23}=39^\circ$, this conclusion is valid for other true $\theta_{23}$ values in the present allowed range. This signifies the fact that,
for a combination of true $\dcp$ and true $\theta_{23}$, where there is no octant degeneracy, addition of antineutrinos from \nova will improve the CP sensitivity but 
for T2K it will reduce the sensitivity.

\section{Combined CP sensitivity of T2K, \nova and ICAL: Replacing the T2K antineutrino run}
\label{sec5}

After discussing the role of antineutrinos in detecting CP violation in T2K and \nova and realizing the fact that T2K antineutrinos help only
in a limited true parameter space let us now see what happens if data from other experiments are combined with T2K.
 
In Fig. \ref{ICAL} we plot the CP sensitivity of T2K for different combinations of neutrino and antineutrino runs, when combined with other experiments.
Here we have taken the (3+3) configuration of \nova and 500 kt-yr exposure of the ICAL detector. The first row correspond to the combined sensitivity
of T2K and \nova. In 2nd and 3rd row we present our results for T2K+ICAL and T2K+\nova+ICAL respectively. For each row, the four panels correspond to the
following combinations of true hierarchy and true $\theta_{23}$: NH-$39^\circ$, NH-$51^\circ$, IH-$39^\circ$ and IH-$51^\circ$. Both hierarchy and octant have been
assumed to be unknown in all the panels. From the first row, we see that
when \nova data is added to T2K, there is an improvement of the overall CP sensitivity. For $\theta_{23} \in$ HO, the hierarchy sensitivity of \nova resolves
the hierarchy-$\dcp$ degeneracy significantly (which occurs at NH-UHP and IH-LHP) as compared to $\theta_{23} \in$ LO. 
But most importantly we notice that for the true combination 
NH-LO-LHP and IH-HO-UHP, there is a significant improvement of the CP sensitivity as compared to the top right panel of Fig. \ref{NH}, when T2K runs in dominant neutrino mode.
We see that T2K(7+1) gives the maximum sensitivity for all the combinations of true parameters when \nova(3+3) is added to T2K. 
This is in stark contrast to the scenario
when T2K alone runs in dominant neutrino mode for this true combinations. 
This is because except for adding CP
sensitivity of its own, the antineutrino component 
of \nova also provides the required sensitivity to lift the wrong octant solutions which appears due to the dominant neutrino runs of T2K. 
In the next row, we see that,
in all the panels the best sensitivity comes from T2K(7+1) when the atmospheric data of ICAL is added to it. As atmospheric neutrinos themselves do not have 
much CP sensitivity by on its own \cite{Ghosh:2013yon,Ghosh:2014dba}, the overall CP sensitivity remains the same but from the figures we see that apart 
from removing the wrong octant solutions in 
NH-LO-LHP and NH-HO-UHP, the hierarchy sensitivity of ICAL also resolves the hierarchy-$\dcp$ degeneracy which appears in NH-UHP and IH-LHP in T2K. 
So here the improvement in the CP sensitivity is two fold.
Finally in the third row we present the combined sensitivity of T2K, \nova and ICAL. In these figures we see that, all possible combinations of true hierarchy and true 
octant are now almost free from both type I and type II degeneracies and there is an improvement of the overall CP sensitivity. It is also clear from the figures that
for every true value, it is possible to have best CP sensitivity from T2K(7+1).

 \section{Summary and Conclusion}
 \label{sec6}
 
In this work we have studied the physics of antineutrino run of T2K for 
detecting CP violation and also studied the possibility to reduce the antineutrino run of T2K 
 by adding the data from other experiments. We have shown that for T2K,
 if octant is known then antineutrinos have no role to play 
 and addition of antineutrinos cause decrease in the CP sensitivity due to lesser statistics. This is true for any true combination of hierarchy, octant and $\dcp$.
 If octant is unknown,
 then we see that T2K antineutrino runs are required only for the true combination of NH-LO-LHP and IH-HO-UHP as these true combinations suffer from
 the octant-$\dcp$ degeneracy which behaves differently for
neutrinos and antineutrinos. We also notice that the CP sensitivity of T2K is also compromised due to the
 hierarchy-$\dcp$ degeneracy for NH-HO and IH-LO and addition of antineutrinos do not help in those regions as this degeneracy is same for
 neutrinos and antineutrinos. We find that this situation is slightly different in \nova. For \nova, antineutrinos help in improving the CP sensitivity even
 when the octant is known. From this we can understand that unlike T2K, 
 \nova antineutrino can help in increasing the CP sensitivity for the true parameter space where octant degeneracy is absent.
 As T2K antineutrino run is useful for only 25$\%$ of the true parameter space and running T2K in equal neutrino-antineutrino mode will affect the CP
 sensitivity in the other $75\%$ of the true parameter space, we ask the question if the T2K antineutrino run can be reduced by adding data from other experiments.
 We find that when the data from \nova(3+3) and and ICAL are added to it, it is indeed possible to obtain the maximum CP sensitivity from the T2K(7+1) configuration. 
 We notice that addition of only \nova(3+3) data to T2K help in removing the wrong octant solutions in the respective true parameter combination and also removes
 the wrong hierarchy solutions significantly for $\theta_{23} \in$ HO. 
 The CP sensitivity of \nova is also added to the CP sensitivity of T2K and causes an overall improvement
 in the CP sensitivity. When the only ICAL data is added to T2K, we find that the overall improvement of CP sensitivity is not that much but the hierarchy and octant
 sensitivity of the atmospheric data resolve both the wrong hierarchy and wrong octant solutions for every values of $\dcp$. We also showed that when both
 \nova and ICAL are added to T2K, we obtain an improved CP sensitivity with almost no degeneracy and the best CP sensitivity can be obtained from the T2K(7+1) combination.
 Though we understand that when T2K will complete
 its run, ICAL data will not be available but it can be noticed that the amount of sensitivity required to solve the degeneracy for the CPV discovery is 
 not too high and it can also come from other atmospheric data which will be available at that time, for example data of Super Kamiokande experiment.
 Though we have only considered the value of $\theta_{23}$ as $39^\circ$ for LO and $51^\circ$ for IH, the physics of the antineutrinos in T2K as discussed in this work
 is true for every value of $\theta_{23}$.
 even those close to maximal value. 
 The results presented in this work are very important for the planning of future antineutrino runs of T2K and in understanding the synergy of 
 T2K with other different experiments.

 \section*{Acknowledgments}
 The author would like to thank Prof. Srubabati Goswami for her encouragement and critical reading of the manuscript. 
 The author would also like to thank Dr. Sushant Raut for
 checking the manuscript thoroughly in WHEPP XIV meeting. A special thanks to Dr. Pomita Ghoshal for her help in learning the atmospheric neutrino code.
 The author also thanks Newton Nath for useful discussion.

\end{document}